\documentclass[a4paper]{article}

\usepackage{icrc2013}

\usepackage{upgreek}

\usepackage{amsmath}
\usepackage[english]{babel}
\title{High-energy cosmic rays measured with KASCADE-Grande}
\newcommand{\etal}{\MakeLowercase{\textit{et al. }}} % "et al."
\shorttitle{Andreas Haungs \etal KASCADE-Grande}

\authors{
A.~Haungs$^{1}$,
W.D.~Apel$^{1}$,
J.C.~Arteaga-Vel\'azquez$^{2}$,
K.~Bekk$^{1}$,
M.~Bertaina$^{3}$,
J.~Bl\"umer$^{1,4}$,
H.~Bozdog$^{1}$,
I.M.~Brancus$^{5}$,
E.~Cantoni$^{3,6,a}$,
A.~Chiavassa$^{3}$,
F.~Cossavella$^{4,b}$,
C.~Curcio$^{3}$,
K.~Daumiller$^{1}$,
V.~de~Souza$^{7}$,
F.~Di~Pierro$^{3}$,
P.~Doll$^{1}$,
R.~Engel$^{1}$,
J.~Engler$^{1}$,
B.~Fuchs$^{4}$,
D.~Fuhrmann$^{8,c}$,
H.J.~Gils$^{1}$,
R.~Glasstetter$^{8}$,
C.~Grupen$^{9}$,
D.~Heck$^{1}$,
J.R.~H\"orandel$^{10}$,
D.~Huber$^{4}$,
T.~Huege$^{1}$,
\mbox{K.-H.}~Kampert$^{8}$,
D.~Kang$^{4}$, 
H.O.~Klages$^{1}$,
K.~Link$^{4}$, 
P.~{\L}uczak$^{11}$,
M.~Ludwig$^{4}$,
H.J.~Mathes$^{1}$,
H.J.~Mayer$^{1}$,
M.~Melissas$^{4}$,
J.~Milke$^{1}$,
B.~Mitrica$^{5}$,
C.~Morello$^{6}$,
J.~Oehlschl\"ager$^{1}$,
S.~Ostapchenko$^{1,d}$,
N.~Palmieri$^{4}$,
M.~Petcu$^{5}$,
T.~Pierog$^{1}$,
H.~Rebel$^{1}$,
M.~Roth$^{1}$,
H.~Schieler$^{1}$,
S.~Schoo$^{1}$,
F.G.~Schr\"oder$^{1}$,
O.~Sima$^{12}$,
G.~Toma$^{5}$,
G.C.~Trinchero$^{6}$,
H.~Ulrich$^{1}$,
A.~Weindl$^{1}$,
D.~Wochele$^{1}$,
J.~Wochele$^{1}$,
J.~Zabierowski$^{11}$ \\
KASCADE-Grande Collaboration
}

\afiliations{
$^1$ Institut f\"ur Kernphysik, KIT - Karlsruher Institut f\"ur Technologie, Germany\\
$^2$ Universidad Michoacana, Instituto de F\'{\i}sica y Matem\'aticas, Morelia, Mexico\\
$^3$ Dipartimento di Fisica, Universit\`a degli Studi di Torino, Italy\\
$^4$ Institut f\"ur Experimentelle Kernphysik, KIT - Karlsruher Institut f\"ur Technologie, Germany\\
$^5$ National Institute of Physics and Nuclear Engineering, Bucharest, Romania\\
$^6$ Osservatorio Astrofisico di Torino, INAF Torino, Italy\\
$^7$ Universidade S$\tilde{a}$o Paulo, Instituto de F\'{\i}sica de S\~ao Carlos, Brasil\\
$^8$ Fachbereich Physik, Universit\"at Wuppertal, Germany\\
$^9$ Department of Physics, Siegen University, Germany\\
$^{10}$ Dept. of Astrophysics, Radboud University Nijmegen, The Netherlands\\
$^{11}$ National Centre for Nuclear Research, Department of Cosmic Ray Physics, Lodz, Poland\\
$^{12}$ Department of Physics, University of Bucharest, Bucharest, Romania\\
\scriptsize{
$^{a}$ now at: Istituto Nazionale di Ricerca Metrologia, INRIM, Torino; \\
$^{b}$ now at: Max-Planck-Institut Physik, M\"unchen, Germany; \\
$^{c}$ now at: University of Duisburg-Essen, Duisburg, Germany;  \\
$^{d}$ now at: University of Trondheim, Norway.
}
}
\email{andreas.haungs@kit.edu}

\abstract{The detection of high-energy cosmic rays above a few hundred TeV is realized by the observation 
of extensive air-showers. By using the multi-detector setup of KASCADE-Grande, energy spectrum, 
elemental composition, and anisotropies of high-energy cosmic rays in the energy range from below the 
knee up to 2 EeV are investigated. In addition, the large high-quality data set permits distinct tests 
of the validity of hadronic interaction models used in interpreting air-shower measurements. 
After more than 16 years, the KASCADE-Grande experiment terminated measurements end of 2012. 
This contribution will give an overview of the main results of the data analysis achieved so far, 
and will report about the status of KCDC, the 'KASCADE Cosmic-ray Data Center', where via a web-based 
interface the data will be made available for the interested public. 
}

\begin{document}
\maketitle

\section{KASCADE-Grande}
Main parts of the experiment are the Grande array spread over an area of $700 \times 700\,$m$^2$, 
the original KASCADE array covering $200 \times 200\,$m$^2$ with unshielded and shielded 
detectors, a large-size hadron calorimeter, and additional muon tracking devices. 
The radio antenna field LOPES~\cite{lopes} and the microwave experiment CROME~\cite{crome} 
with all its components complete the experimental set-up of KASCADE-Grande 
at the Karlsruhe Institute of Technology in Germany.
This multi-detector system allows us to 
investigate the energy spectrum, composition, and anisotropies of cosmic rays in the energy 
range up to and even above $1\,$EeV. In addition, the data is used for detailed cross-checks of the 
hadronic interaction models underlying the analysis of extensive air shower reconstruction 
as well as basis for R\&D-studies of the radio detection technique.

KASCADE-Grande stopped finally the active data acquisition of all its components end of 
2012 and is presently being decommissioned. The collaboration, however, continues the detailed 
analysis of nearly 20 years of data. Moreover, with KCDC, the KASCADE Cosmic-ray Data Center,
we plan to provide to the public the edited data via a customized web page.  

\begin{figure}[!t]
  \vspace{5mm}
  \centering
  \includegraphics[width=0.75\columnwidth]{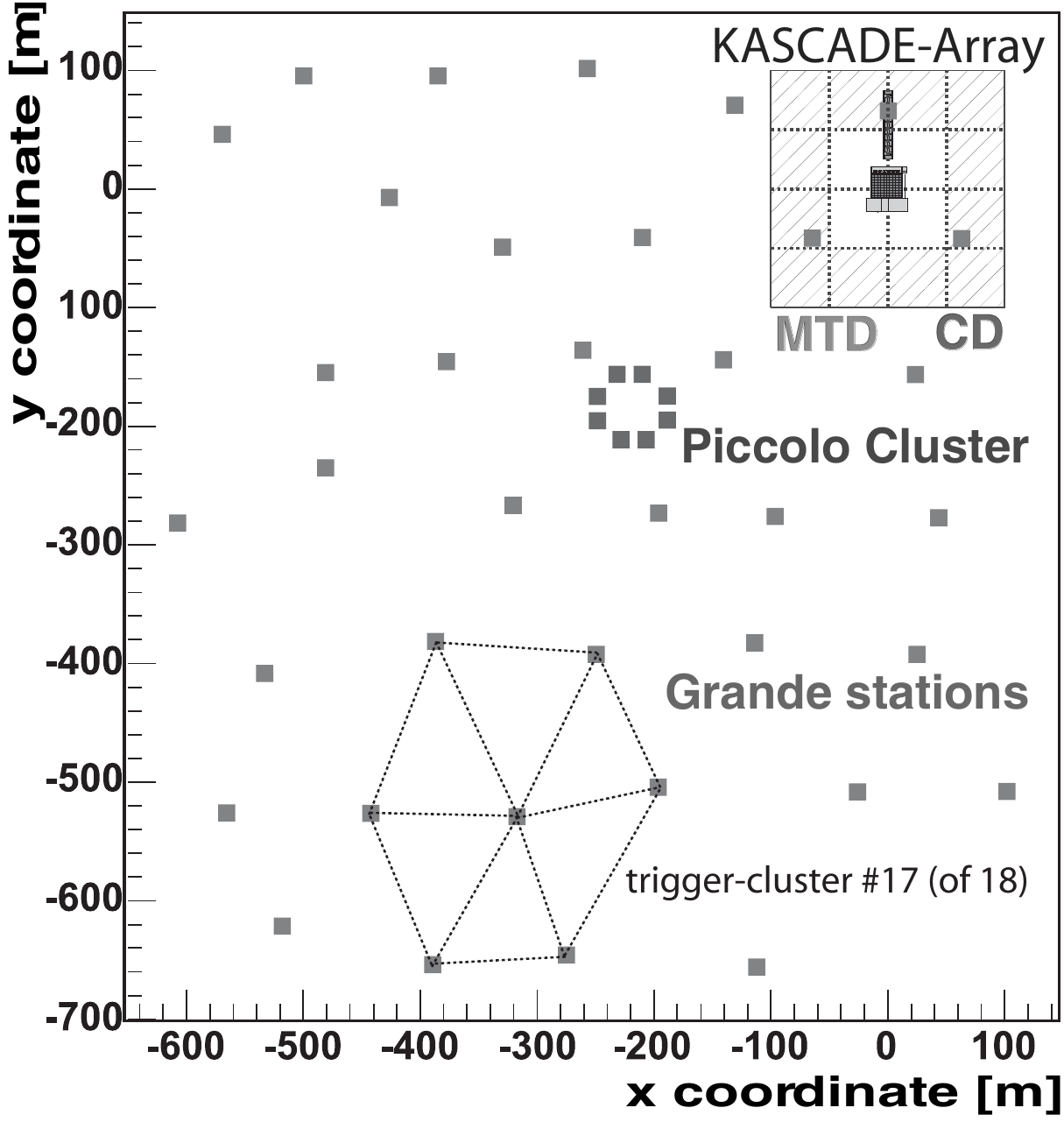}
	\caption{Layout of the KASCADE-Grande experiment: The original KASCADE, the distribution of 
	the 37 stations of the Grande array, and the small Piccolo cluster for fast trigger 
	purposes are shown. The outer 12 clusters of the KASCADE array consist of $\mu$- and $e/\gamma$-detectors, 
	the inner 4 clusters of $e/\gamma$-detectors, only. }
	\label{fig1}
 \end{figure}

The estimation of energy and mass of the primary particles is based 
on the combined investigation of the charged particle, the electron, and the muon components 
measured by the detector arrays of Grande and KASCADE. 
The multi-detector experiment KASCADE~\cite{kascade}
(located at 49.1$^\circ$n, 8.4$^\circ$e, 110$\,$m$\,$a.s.l.)
was extended to KASCADE-Grande 
in 2003 by installing a large array of 37 stations consisting 
of 10$\,$m$^2$ scintillation detectors each (fig.~\ref{fig1}).  
KASCADE-Grande~\cite{kg-NIM10} provides an area of 0.5$\,$km$^2$
and operates jointly with the existing KASCADE detectors.
The joint measurements with the KASCADE muon tracking devices were 
ensured by an additional cluster (Piccolo) 
close to the center of KASCADE-Grande for fast trigger purposes. 
While the Grande detectors are sensitive to charged particles, 
the KASCADE array detectors measure the electromagnetic 
component and the muonic component separately. 
These muon detectors enable to reconstruct 
the total number of muons on an event-by-event basis
also for Grande triggered events. 
The strategy of the KASCADE-Grande data analysis to reconstruct the energy spectrum and 
elemental composition of cosmic rays is to use the multi-detector set-up of the experiment and 
to apply different analysis methods to the same data sample. This has advantages in various 
aspects: One would expect the same results by all 
methods when the measurements are accurate enough, when the reconstructions work 
without failures, and when the Monte-Carlo simulations describe correctly and consistently the 
shower development and detector response. 

\section{The all-particle energy spectrum}
In a first step of the Grande data analysis, we reconstructed the all-particle energy spectrum.
By combining both observables and using the hadronic interaction model QGSJet-II, a composition
independent all-particle energy spectrum of cosmic rays is reconstructed in the energy range 
of $10^{16}\,$eV to $10^{18}\,$eV within a total uncertainty in flux of 10-15\%~\cite{Apel2012183}.

\begin{figure}[!t]
   \centering
  \includegraphics[width=0.9\columnwidth]{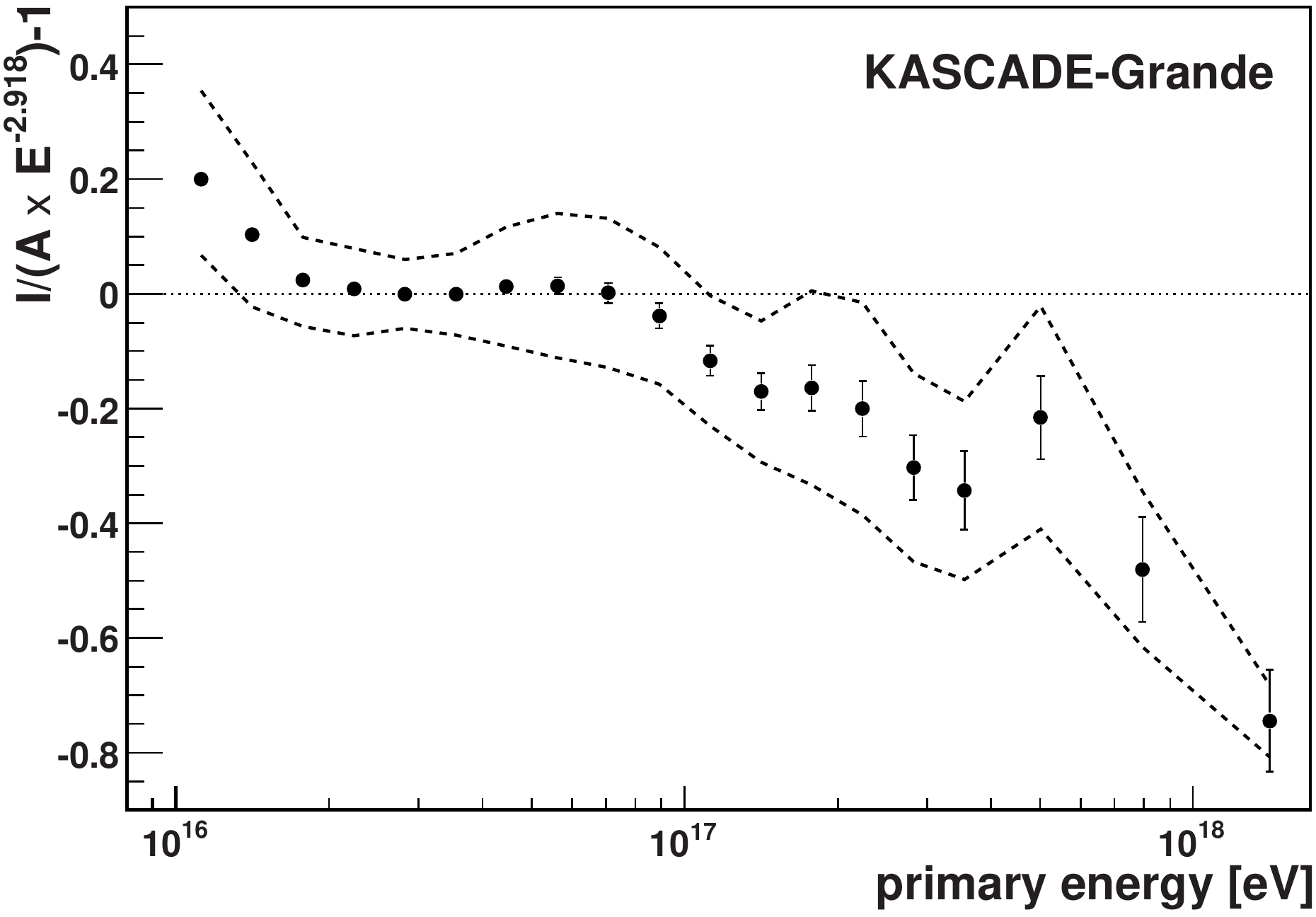}
	\caption{The all-particle energy spectrum obtained with KASCADE-Grande based on the 
	QGSJet-II model and unfolded, i.e. corrected for the reconstruction uncertainties. 
	Shown is the residual flux as well as the band of systematic uncertainty~\cite{Apel2012183}.}
	\label{residual}
 \end{figure}

Despite the overall smooth power law behavior of the resulting all-particle spectrum, 
there are some structures observed, which do not allow to describe the spectrum with a single 
slope index. Figure~\ref{residual} shows the resulting 
all-particle energy spectrum multiplied with a factor in such a way 
that the middle part of the spectrum becomes flat. 
There is a clear evidence that just above $10^{16}\,$eV the spectrum shows a `concave' behavior, 
which is significant with respect to the systematic and statistical uncertainties.   
Another feature in the spectrum is a small break, i.e. knee-like feature at around $10^{17}\,$eV. 
This slight slope change occurs at an energy where the rigidity dependent knee 
of the iron component would be expected. 
Despite the fact, that the discussed spectrum is based on the QGSJet-II hadronic interaction model, 
there is confidence that the found structures of the energy spectrum remain stable~\cite{marioJASR}. 

\section{Composition}
\begin{figure}
\includegraphics[width=\columnwidth]{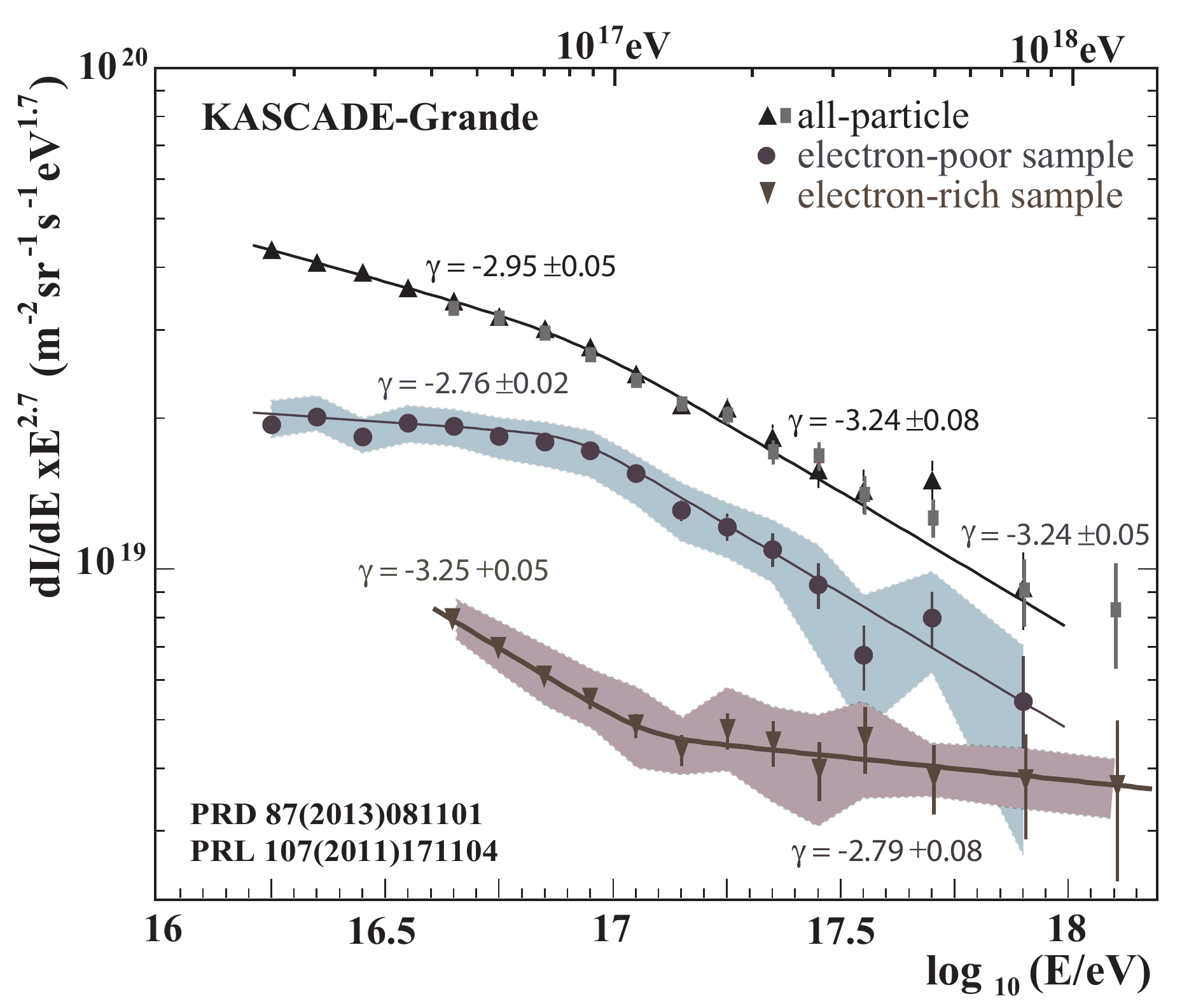}
\caption{All-particle, electron-poor, and electron rich energy spectra from KASCADE-Grande. The
all-particle and heavy enriched spectrum is from ref.~\cite{prl107} and the all-particle and light 
primary spectrum results from a larger data set with higher energy threshold described in ref.~\cite{Apel2013}.}
\label{3spectra}
\end{figure}

The basic goal of the KASCADE-Grande experiment is the determination of the chemical 
composition in the primary energy range $10^{16} - 10^{18}\,$eV by reconstructing 
individual mass group spectra.  

The main observables taken into account for the recent composition studies at 
KASCADE-Grande are the shower size $N_{ch}$ and the muon shower size $N_\mu$. 
Using the above mentioned reconstruction of the energy spectrum by correlating 
$N_{ch}$ and $N_{\mu}$ on an event-by-event basis, the mass 
sensitivity is minimized by means of a parameter $k(N_{ch},N_{\mu})$. 
On the other hand, the evolution of $k$ as a function of energy keeps track of the 
evolution of the composition, and allows an event-by-event separation between light, 
medium and heavy primaries, at least. 
Using $k$ as separation parameter for different mass groups, where the values 
of $k$ have to be determined with help of simulations, directly the energy spectra of 
the mass groups are obtained~\cite{prl107,Apel2013}. 
The application of this methodical approach to shower selection and separation in various 
mass groups were performed and cross-checked in different ways, where figure~\ref{3spectra} 
shows the main results:

{\bf Knee-like feature in the heavy component of primary cosmic rays:} 
The reconstructed spectrum of the electron-poor events, i.e. the spectrum of heavy primaries, shows 
a distinct knee-like feature at about $8 \cdot 10^{16}\,$eV. 
Applying a fit of two power laws to the spectrum interconnected by a smooth 
knee results in a statistical significance of $3.5\sigma$ that the 
entire spectrum cannot be fitted with a single power law.
The change of the spectral slope is $\Delta \gamma = -0.48$ from
$\gamma = -2.76 \pm 0.02$  to $\gamma = -3.24\pm0.05$ with the break position at
$\log_{10}(E/eV)=16.92\pm0.04$. 
Applying the same function to the all-particle spectrum results in a statistical
significance of only $2.1\sigma$ at the same energy and a change of the spectral slope 
from $\gamma = -2.95 \pm 0.05$  to $\gamma = -3.24\pm0.08$. 
Hence, the selection of heavy primaries enhances the knee-like feature that is already 
present in the all-particle spectrum. 
\begin{figure}[!t]
\includegraphics[width=\columnwidth]{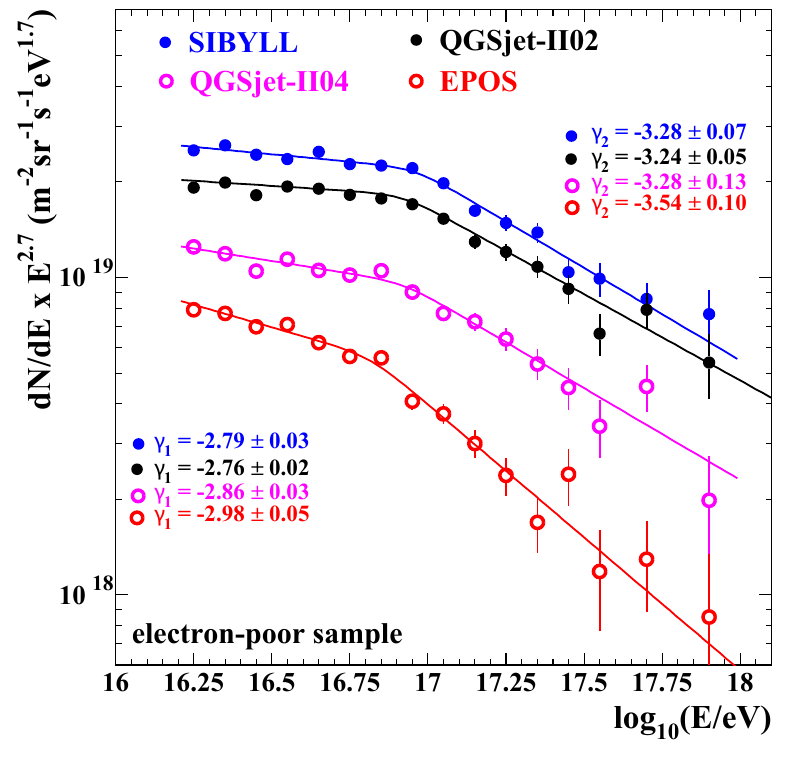}
\caption{Reconstructed energy spectra of the heavy primary component for four hadronic interaction models. 
The error bars show the statistical uncertainties; fits on the spectra and resulting slopes 
before and after the heavy knee are also indicated~\cite{icrcmario}}.
\label{heavymodels}
\end{figure}

{\bf Ankle-like feature in the light component of primary cosmic rays:} 
An ankle-like feature is clearly visible in the spectrum of the electron-rich events, e.g. 
light elements of the primary cosmic rays at an energy of  $10^{17.08 \pm 0.08} \, \mathrm{eV}$. 
At this energy, the spectral index changes by $\Delta \gamma = 0.46$
from $\gamma_{1} = -3.25 \pm 0.05$ to $\gamma_{2} = -2.79 \pm 0.08$.
Applying again a fit of two power laws to the spectrum results in a statistical significance 
of $5.8\sigma$ that the entire spectrum cannot be fitted with a single power law.
It is worth to mention that the changes in the spectrum of heavy primaries and in the spectrum of light 
elements are not connected by a bias in the separation or reconstruction procedures, which was checked 
in detail~\cite{icrcsven,icrckang,icrcmario}.

It is crucial to verify the sensitivity of the observables to different primary 
particles and the reproducibility of the measurements with the hadronic interaction model in use as 
a function of sizes and the atmospheric depth. 
As it is well known from KASCADE data analysis~\cite{kas-unf} that the relative abundances of the 
individual elements or elemental groups are very dependent on the hadronic interaction model underlying 
the analyses, the strategy is to derive the energy spectra of the individual mass groups by applying 
different methods of composition analysis (see also~\cite{icrcandrea,icrcdaniel}) 
and on basis of different hadronic interaction models~\cite{marioJASR,icrckang,icrcmario}. 
Cross-checks of the general validity of the hadronic interaction models are performed within 
KASCADE-Grande by detailed investigations of the muon component (first results see~\cite{icrcjuan,icrcpawel}). 
The structure or characteristics of the spectra are found to be much less affected by the differences of 
the various hadronic interaction models than the relative abundances of the masses. An example is shown 
in figure~\ref{heavymodels} where the electron-poor spectra are displayed for the reconstruction on basis of 
four different hadronic interaction models. 

\section{The KASCADE Cosmic-ray Data Center KCDC}
\begin{figure}
\includegraphics[width=\columnwidth]{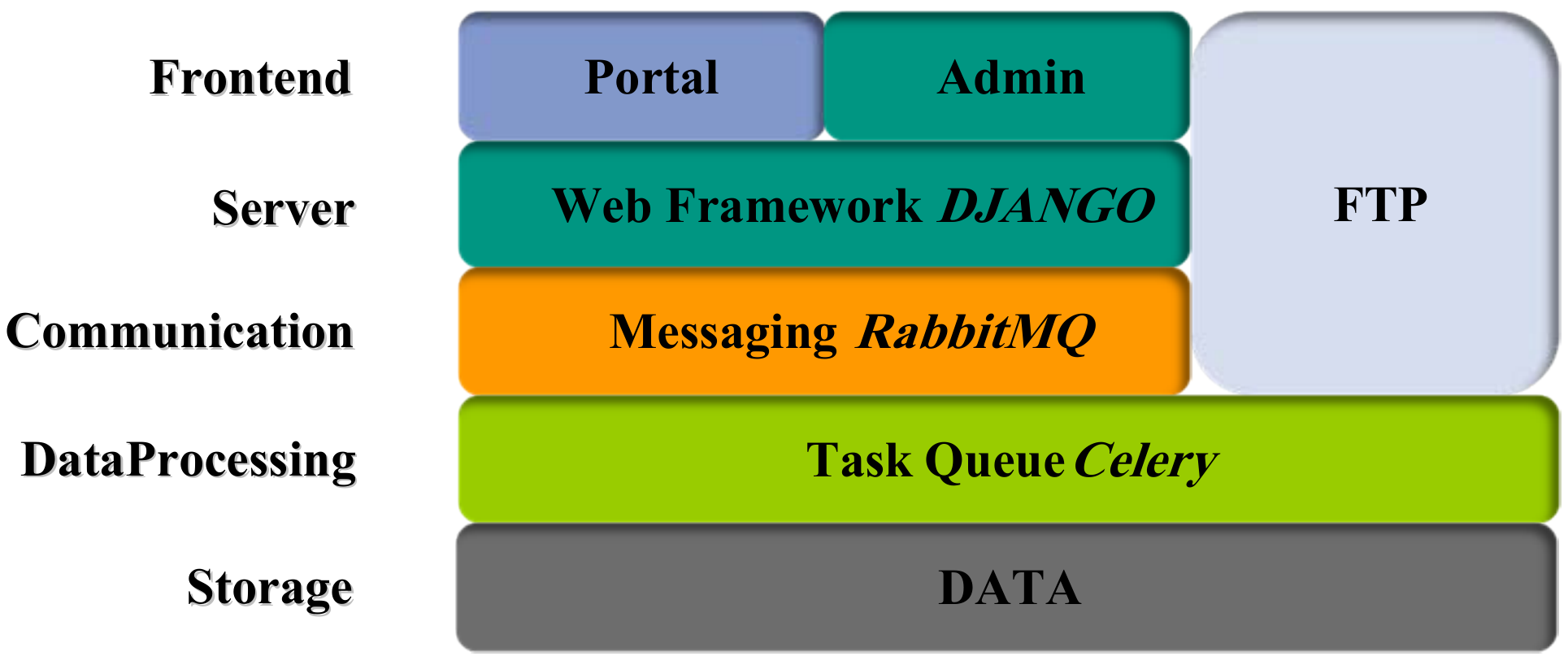}
\caption{Software structure of KCDC, the KASCADE Cosmic-ray Data Center}
\label{KCDCsoft}
\end{figure}
The KASCADE/KASCADE-Grande experiment was a 
large-area detector for the measurement of cosmic ray air showers lasting for more than 20 years and 
financed by taxes. 
The aim of this particular project is the installation and establishment of a public 
data center for high-energy astroparticle physics. 
In the research field of astroparticle physics, such a data release is a novelty, whereas the data 
publication in astronomy has been established for a long time. 
Therefore, there are no completed concepts, how the data can be treated/processed so that they are 
reasonably usable outside the collaboration.
The first goal of KCDC is to make to the community the data from the KASCADE 
experiment available. 
A concept for this kind of data center (software and hardware) is already developed (fig.~\ref{KCDCsoft}) 
and implemented, and will soon be opened as a platform to external users. 
The project faces thereby open questions, e.g. how to ensure a consistent calibration, 
how to deal with data filtering and how to provide the data in a portable format as well as how a
sustainable storage solution can be implemented. 
In addition, access rights and license policy play a major role and have to be considered.
KCDC is foreseen to be released in  2013.

\section{Summary}
 \begin{figure*}[t]
   \centering
  \includegraphics[width=0.83\textwidth]{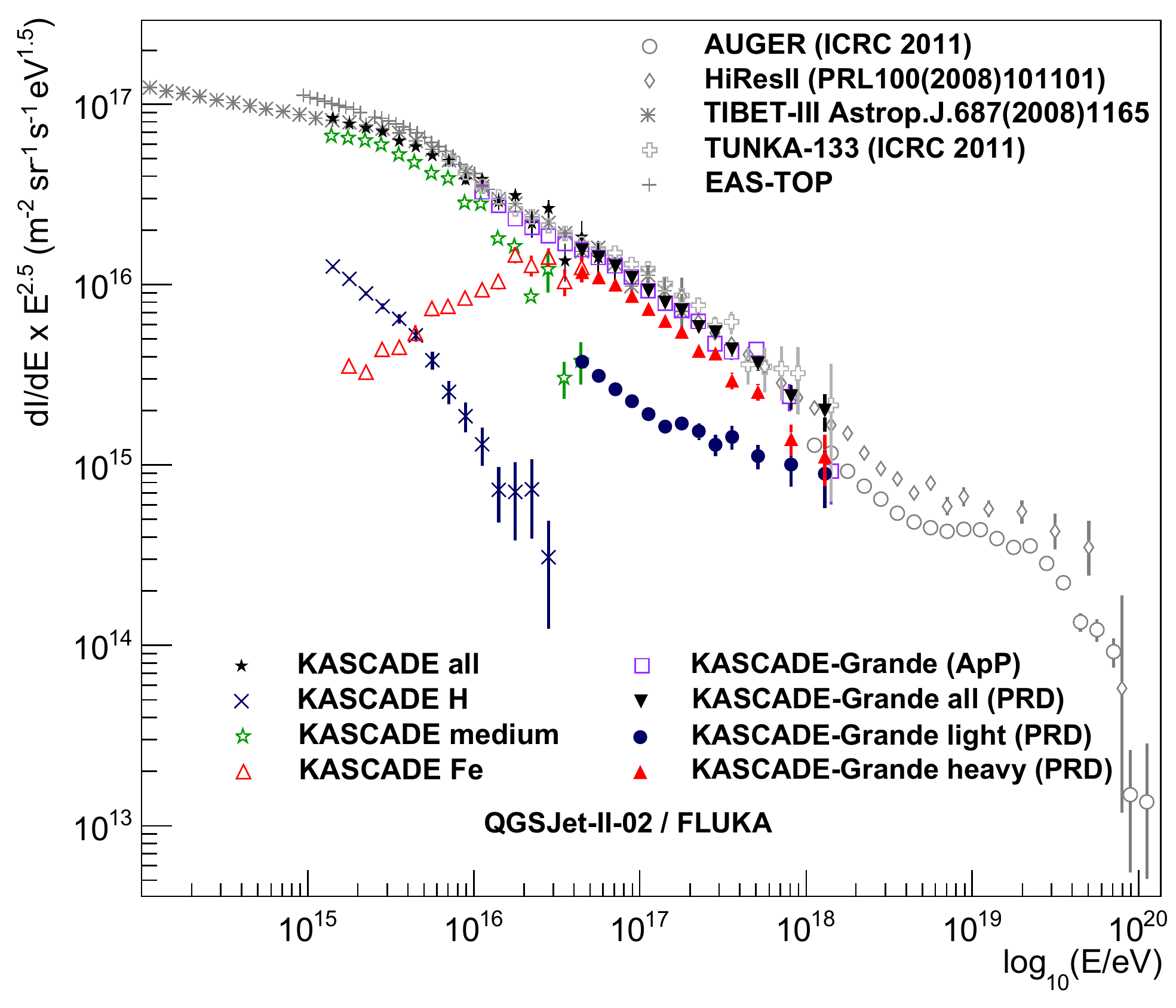}
	\caption{The all-particle energy spectrum and spectra of individual mass groups obtained with 
	KASCADE and KASCADE-Grande.}
	\label{allspec}
 \end{figure*}
In summary, after separating the KASCADE-Grande measured events into a light and a heavy component, 
a knee-like feature is identified in the spectrum of the heavy component, and an ankle-like feature 
is observed in the spectrum of the light component. 
Whereas the 'heavy-knee' occurs at an energy where the rigidity dependent knee of the iron component 
is expected, the 'light-ankle' might indicate an early transition from galactic to extragalactic 
origin of cosmic rays. Figure~\ref{allspec} show the situation of the cosmic-ray energy spectrum 
in a wide energy range taking into account the results from KASCADE and KASCADE-Grande.

\section{Acknowledgments}
KASCADE-Grande is supported by the BMBF and the 'Helmholtz Alliance for Astroparticle Physics - HAP'
of Germany, the MIUR and INAF of Italy, the Polish Ministry of Science and Higher Education and the 
Romanian Authority for Scientific Research(UEFISCDI (PNII-IDEI grants 271/2011 and 17/2011).

\clearpage

\end{document}